\documentclass[twocolumn,showpacs,preprintnumbers]{revtex4}%
\usepackage{amsmath}
\usepackage{amsfonts}
\usepackage{amssymb}
\usepackage{graphicx}%
\providecommand{\U}[1]{\protect\rule{.1in}{.1in}}
\begin{document}
\title{From the Hosftadter to the Fibonacci butterfly}
\author{Gerardo G. Naumis, F.J. L\'{o}pez-Rodr\'{\i}guez}
\affiliation{Instituto de F\'{\i}sica, Universidad Nacional Aut\'{o}noma de M\'{e}xico
(UNAM), Apartado Postal 20-364, 01000, M\'{e}xico, Distrito Federal, Mexico.}
\date{\today }

\begin{abstract}
We show that the electronic spectrum of a tigth-binding hamiltonian defined in
a quasiperiodic chain with an on-site potential given by a Fibonacci sequence,
can be obtained as a superposition of Harper potentials. The electronic
spectrum of the Harper equation is a fractal set, known as Hosfateedter
butterfly. Here we show that is possible to construct a similar butterfly for
the Fibonacci potential just by adding harmonics to the Harper potential. As a
result, the equations in reciprocal space for the Fibonacci case have the form
of a chain with a long range interaction between Fourier components. Then we
explore the transformation between both spectra, and specially the origin of
energy gaps due to the analytical calculation of the components in reciprocal
space of the potentials. We also calculate some localization properties by
finding the correlator of each potential.

\end{abstract}

\maketitle

\section{Introduction}

Although the discovery of quasicrystals \cite{Blech}, which are alloys with an
structure neither periodic, nor disordered, launched an extensive
investigation on quasiperiodic Hamiltonians, in fact the study of such
Hamiltonians goes back to the old Frenkel-Kontorova model \cite{Bambi} and to
the pioneer research made in the seventies \cite{Hofstadter,Aubry}. Here the
word quasiperiodic means that in the system there are incommensurate periods;
and as a result, the dimension of the Fourier space is always bigger than the
dimensionality of the system. One of the most famous quasiperiodic
Hamiltonians was obtained by Harper in connection with a problem proposed by
Peierls \cite{Hofstadter}. The idea was to find the spectrum and the
wave-functions of an electron in a square lattice with a perpendicular
magnetic field. Two periods are involved in the problem, the electron motion
in the lattice and the cyclotron frequency \cite{Hofstadter}. The spectrum as
a function of ratio between these periods turned out to be a complex set known
as the Hostaedter butterfly \cite{Hofstadter} (see figure 1). Since then, the
Harper model has been very useful to investigate the transition from localized
to extended eigenstates, as the spectrum pass from pure point to continuous
\cite{Hofstadter}\cite{Aubry}\cite{Ketmerick}. Between both limits, there is a
new type of spectrum which is known as singular continuous \cite{Aubry}. The
corresponding eigenstates are called critical and show self-similar
properties. For certain parameters of the Harper equation, the distribution of
level spacings follows an inverse power law \cite{Machida}, which is a new
type of spectral statistics \cite{Machida}, explained as a level clustering
tendency \cite{Geisel}. It has been possible to find analytical expressions
for the wave-functions using quantum groups \cite{Wiegmann}. More recently,
the quantum phase diagrams \cite{Koshino} and the electronic correlation
effects have been analyzed \cite{Czajka}.

Another quasiperiodic system that has been extensively studied is the
Fibonacci chain (FC). This chain is the simplest model of a quasicrystal
\cite{Kohmoto}. The importance of this Hamiltonian arises because the nature
of the physical properties of quasicrystals is still not well understood
\cite{Thiel}\cite{Vedemenko}\cite{Moras}. Even in the theoretical side there
is a lack of understanding in how electrons propagate, specially in two and
three dimensions \cite{Maciarep}. As is well known, a periodic potential
satisfies the Bloch%
%TCIMACRO{\U{b4}}%
%BeginExpansion
\'{}%
%EndExpansion
s theorem, which tells that the eigenstates of the Schr\"{o}edinger equation
are plane waves of delocalized nature, and the energy spectrum is continuous
\cite{Kittel}. For disordered systems, like in the one dimensional (1D)
Anderson model, all the states are localized corresponding to isolated
eigenvalues \cite{Zallen}. In more dimensions, there is a mobility edge which
separates extended from localized states \cite{Zallen}. For most of the
quasiperiodic systems in 1D, the spectrum is neither continuous nor singular,
instead a new type of spectrum, called singular continuous is obtained
\cite{Suto}. This kind of spectrum is similar to a Cantor set, and presents a
multifractal nature. The corresponding eigenfunctions are called critical, and
also show self-similarity and multifractality. In two and three dimensions,
the nature of the spectrum is not known, although there seems to be a kind of
mobility edge \cite{NaumisPenrose}\cite{NaumisJPC}. However, even in 1D, where
large amount of work has been done, there are many unsolved questions, like
the nature of conductivity \cite{Dominguez} or diffusivity \cite{Ketmerick},
the spectral statistics and the shape of many of the eigenfunctions
\cite{Fujiwara}\cite{Macia}. Even in the FC, there are no analytical
expression for the wave functions, except for few energies \cite{Kohmoto}.

The Harper and Fibonacci potentials share many characteristics; for example,
both can present a multifractal spectrum with self-similar wave functions,
although the Harper equation can also present pure point and continuous
spectrum. An interesting question is why Fibonacci does not show pure point or
continuous spectrum. Furthermore, is it possible to build the analogous of the
Hostadter butterfly for the FC? An understanding of these questions will serve
to give a better picture of the electronic properties of quasiperiodic
systems. For example, it can suggest a way to construct analytical solutions
for the FC in terms of those solutions from Harper.

In this article, we show that in fact, the Fibonacci potential can be made
from a superposition of Harper potentials. Then, we can follow the transition
of the Hofstadter butterfly to the equivalent in the Fibonacci case. This
allows to explore the equations in reciprocal space of the FC. The layout of
this work is the following, in section II we show how to obtain the Fibonacci
potential in terms of Harper. Section III is devoted to a discussion of the
corresponding spectra using the properties in reciprocal space, while section
IV is devoted to study the localization in terms of the correlators of the
potential. Finally, in section V the conclusions are given.

\section{The Fibonacci and Harper models}

As a general model we will use a tight-binding Hamiltionian of the type,%

\begin{equation}
\left(  E-V(n)\right)  \psi_{n}=t_{n}\psi_{n+1}+t_{n-1}\psi_{n-1},\label{tb}%
\end{equation}
where $\psi_{n}$ is the wave-function at site $n$, $t_{n}$ is the resonance
integral between sites $n$ and $n+1$. For the present purposes, $t_{n}$ is set
to $1$ for all sites. $V(n)$ is the atomic on-site potential and $\ E$ are the
allowed energies. The generalized Harper equation is obtained when
\cite{Hofstadter},
\begin{equation}
V(n)\equiv V_{H}(n)\equiv2\lambda\cos(2\pi\phi n+\upsilon),\label{Harperpot}%
\end{equation}
where $\lambda\geq0$ is the strength of the potential, $\phi$ is a parameter
that contains the ratio between the electron cyclotron frequency and the
elementary quantum flux, and $\upsilon$ is a phase shift. For a rational
$\phi$, Eq. (\ref{tb}) can be solved by Bloch's theorem since the problem is
periodic, although the value of such theorem is very limited since the
coefficients in Fourier space of the solution form a dense set
\cite{Hofstadter}. For $\phi$ irrational, the spectrum depends on the value of
$\lambda$. For $\lambda<1,$ the spectrum is continuous with extended wave
functions, for $\lambda>1$ the spectrum is made from pure points and localized
solutions. For $\lambda=1$ the spectrum is singular continuous with
self-similar wave functions.

The other potential that we will consider, is the simplest model of a
quasicrystal, obtained when $V(n)$ has two possible values, $V_{A}$ and
$V_{B}$ following the Fibonacci sequence (FS). The FS is build as follows:
consider two letters, $A$ and $B$, and the substitution rules, $A\rightarrow
B,$ and $B\rightarrow AB.$ If one defines the first generation sequence as
$\mathcal{F}_{1}=A$ and the second one as $\mathcal{F}_{2}=BA$, the subsequent
chains are generated using the two previous rules, for instance,
$\mathcal{F}_{3}=ABA.$ Starting with an $A$, we construct the following
sequences, $A,$ $B$,$AB$, $BAB,$ $ABBAB,$ $BABABBAB,$ and so on. Each
generation obtained by iteration of the rules is labeled with an index $l.$ Is
clear that the number of letters in each generation $l$ is given by the
Fibonacci numbers $F(l)$ of generation $l$, which satisfy:
$F(l)=F(l-1)+F(l-2)$ with the initial conditions: $F(0)=1,F(1)=1.$A Fibonacci
potential is obtained with two values $V_{A},V_{B}$ which follow a FS.

Our first task in order to compare the FC and the Harper potential, consists
in finding an analytical expression for the Fibonacci potential. This can be
done in the following way. By using the cut and projection method, it is very
easy to prove that a one dimensional chain with atoms at positions $y_{n}$
determined by a FS are given by \cite{Naumisphason},%
\[
y_{n}=\left\lfloor n\phi\right\rfloor
\]
where the function $\left\lfloor x\right\rfloor $ denotes the greatest integer
lower than $x,\phi$ \ is a parameter which turns out to be the inverse of the
golden mean $\tau^{-1}=(\sqrt{5}-1)/2$. The separation between atoms is
$y_{n+1}-y_{n}.$ In the Fibonacci sequence, the separation takes two values,
$l_{A}$ and $l_{B}.$ Thus, we can make a Fibonacci potential with the sequence
of spacings to get,%
\begin{equation}
V(n)=V_{B}+V_{A}\left(  \left\lfloor (n+1)\phi\right\rfloor -\left\lfloor
n\phi\right\rfloor \right)  ,
\end{equation}
Using the identity $x=\left\lfloor x\right\rfloor +\{x\},$ where $\{x\}$ is
the decimal part of $x$, we obtain that $V(n)$ can be written as,%
\begin{equation}
V(n)=\left\langle V\right\rangle +\delta V\left(  \{n\phi\}-\{(n+1)\phi
\}\right)  , \label{Fibo}%
\end{equation}
where $\left\langle V\right\rangle =V_{A}\phi+V_{B}(1-\phi)$ is an average
potential that shifts the zero of the energies, and $\delta V$ is the strength
of the quasiperiodicity, measured by the difference between site-energies
$\delta V=V_{A}-V_{B}$. In what follows, without any loss of generality we set
$V_{A}$ and $V_{B}$ in such a way that $\left\langle V\right\rangle =0.$ Now,
if the integer variable $n$ is replaced by a continuous one, say again $x$,
the resulting potential $V(x)$ is just a square wave as shown in figure 1. The
potential stays at -$\delta V\phi$ for an interval of length $1-\phi,$ and
then it jumps to $\delta V(1-\phi)$ for a length $\phi.$

The decimal part function $\{x\}$ has period $1,$ and can be developed as a
Fourier series,%
\[
\{x\phi\}=\frac{1}{2}-\frac{1}{\pi}\sum_{s=1}^{\infty}\frac{1}{s}\sin(2\pi\phi
sx),
\]
It follows that,
\begin{equation}
V(n)=\overline{V}+2\delta V%
%TCIMACRO{\dsum \limits_{s=1}^{\infty}}%
%BeginExpansion
{\displaystyle\sum\limits_{s=1}^{\infty}}
%EndExpansion
\widetilde{V}(s)\cos\left(  \pi s\phi(2n+1)\right)  , \label{pot0}%
\end{equation}
where $\widetilde{V}(s)$ is the $s$ harmonic of the Fourier series,
$\widetilde{V}(s)=\sin(\pi s\phi)/\pi s.$ The first terms of this series are
shown in figure 1, and in fact, we are approximating a square wave by a sum of
cosines. Notice the very slow convergence of the series due to the $1/s$
factor of each harmonic. This potential can be further reduced if a proper
phase $\chi$ is used in Eq. (\ref{Fibo}), in such a way that the terms
$\{x\phi\}$ are replaced by $\{x\phi+\chi\}$. This phase is only a horizontal
translation of the potential. For $\chi=-\phi/2,$ the Fibonacci potential is
simply written as,%

\begin{equation}
V(n)=2\delta V%
%TCIMACRO{\dsum \limits_{s=1}^{\infty}}%
%BeginExpansion
{\displaystyle\sum\limits_{s=1}^{\infty}}
%EndExpansion
\widetilde{V}(s)\cos\left(  2\pi s\phi n\right)  , \label{pot}%
\end{equation}

It is also worthwhile mentioning that for $s=l\phi^{-1}$ (where $l$ is an
integer), $\widetilde{V}(s)=0.$ This condition can only be hold when $\phi$ is
a rational. When this is not the case, $\widetilde{V}(s)\approx0$ for integers
$s$ an $l$ such that $\phi\approx l/s,$ and thus $l$/$s$ is a rational
approximant of $\phi$. Such rationals are obtained from the continuous
fraction development of the number $\phi.$ For the FS, $s\approx l\tau$ from
where it follows that $l$ and $s$ are successive Fibonacci numbers. Such
condition can be interpreted as the existence of a rational approximant
structure when $\phi$ is set to $\phi=l/s$. The most important Fourier
components in Eq. (\ref{pot}), are those harmonics for which $s\approx
(r+1/2)\phi^{-1}$ for an integer $r$. Using the decomposition in integer and
decimal parts, this happens whenever $\{(r+1/2)\phi^{-1}\}$ is nearly zero or one.%

%TCIMACRO{\FRAME{ftbpFU}{2.9974in}{2.1439in}{0pt}{\Qcb{ The Fibonacci potential
%can be obtained by evaluating a square wave of period $\tau$ at integers
%values. The approximation with one, two, three and four harmonics, obtained
%from Eq. (\ref{pot}), are shown in the figure.}}{}{harperfig1.eps}%
%width 2.9974in;  height 2.1439in;  depth 0pt;  original-width 7.7089in;
%original-height 5.5089in;  cropleft "0";  croptop "1";  cropright "1";
%cropbottom "0";  filename 'Harperfig1.eps';file-properties "XNPEU";}}}%
%BeginExpansion
\begin{figure}
[ptb]
\begin{center}
\includegraphics[
height=2.1439in,
width=2.9974in
]%
{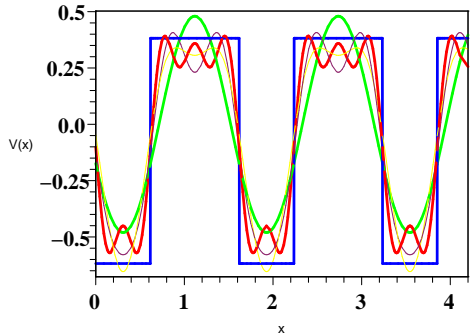}%
\caption{ The Fibonacci potential can be obtained by evaluating a square wave
of period $\tau$ at integers values. The approximation with one, two, three
and four harmonics, obtained from Eq. (\ref{pot}), are shown in the figure.}%
\end{center}
\end{figure}
%EndExpansion

Comparing eq.(\ref{Harperpot}) with eq.(\ref{pot0}) and eq.(\ref{pot}), we
observe that in fact, the Fibonacci potential is just a superposition of
Harper potentials with the right amplitude and phase. In some sense, this is a
similar situation to an applied effective modulated magnetic field
\cite{Yeong}. This leads to many questions. The first is how the transition
from Harper is done. To answer this, we will cut the sum in Eq. (\ref{pot}) at
a finite number of harmonics, denoted by $S$. In figure 2 a) we plot the
energy spectrum of the Harper equation in the case $\lambda=1$. This spectrum,
as well as the others discussed in this article, were obtained by using the
transfer matrix formalism \cite{NaumisJPCM}. Figure 1 a) is the well known
Hosftadter butterfly \cite{Hofstadter}. Notice that here $\upsilon=\pi\phi$,
and thus our figure is not exactly the same as in Hosftadter work, since
therein, a sweep for all values of $\upsilon$ between $0$ and $2\pi$ is made,
and thus the spectrum has more points.%
%TCIMACRO{\FRAME{ftbpFU}{3.6115in}{2.5434in}{0pt}{\Qcb{The energy spectrum as a
%function of the parameter $\phi$ for $\delta V=\pi/\sin(\pi\tau^{-1}%
%),$corresponding to $\lambda=1$ in the pure Harper equation , using a) one
%harmonic (Hofstadter butterfly), b) two harmonics, c) three harmonics, and d)
%the full Fibonacci potential ("Fibonacci butterfly").}}{}{harperfig2com.eps}%
%width 3.6115in;  height 2.5434in;  depth 0pt;  original-width 9.8623in;
%original-height 6.9349in;  cropleft "0";  croptop "1";  cropright "1";
%cropbottom "0";  filename 'Harperfig2com.eps';file-properties "XNPEU";}}}%
%BeginExpansion
\begin{figure}
[ptb]
\begin{center}
\includegraphics[
height=2.5434in,
width=3.6115in
]%
{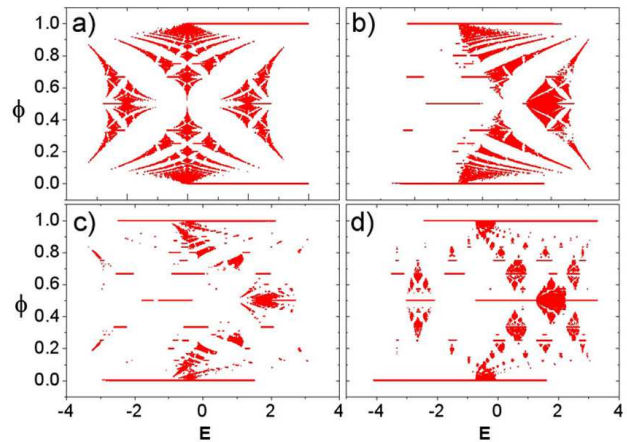}%
\caption{The energy spectrum as a function of the parameter $\phi$ for $\delta
V=\pi/\sin(\pi\tau^{-1}),$corresponding to $\lambda=1$ in the pure Harper
equation , using a) one harmonic (Hofstadter butterfly), b) two harmonics, c)
three harmonics, and d) the full Fibonacci potential ("Fibonacci butterfly").}%
\end{center}
\end{figure}
%EndExpansion

If $\phi$ is taken as a parameter in eq.(\ref{pot}), then the spectrum of Fig.
2a) is also the same as for the Fibonacci potential but with only one
harmonic, $s=1$. The parameter used is $\delta V=\pi/\sin(\pi\tau^{-1}%
)\approx3.3706,$ chosen to correspond to $\lambda=1$ in the Harper equation
$.$ In Figure 2b) and 2c) we show the effects of adding harmonics $s=2$ and
$s=3$ in the development of the potential, and figure 2d) presents the result
for the exact FC. There is an important change between the pure Harper case
and the second harmonic case. Mainly the left part of the Hofstadter butterfly
is washed out. Also, is clear that with only three harmonics, the structure is
already very similar to the pure Fibonacci case. This spectacular wash out due
to the second harmonic has its origins in the difference of lengths in the
steps of the square wave Fibonacci potential. For the first harmonic, this
change is not observed. But by looking at Fig. 1, one can observe that with
two harmonics there is an asymmetry in the upper part of the series. This
effect is more notorious when $\delta V>1$, and can be translated in an almost
split band limit when $\delta V\rightarrow\infty$, around self-energies
$V_{A}$ and $V_{B}$. In fact, this is one of the main differences between the
Harper and Fibonacci potentials.%

%TCIMACRO{\FRAME{ftbpFU}{3.186in}{2.239in}{0pt}{\Qcb{The energy spectrum as a
%function of the parameter $\phi$ for $\delta V=1$ using a) one harmonic
%(corresponding to the Harper model at $\lambda=0.2967)$, b) two harmonics, c)
%three harmonics, and d) full Fibonacci potential.}}{}{harperfig3com.eps}%
%width 3.186in;  height 2.239in;  depth 0pt;  original-width 10.3639in;
%original-height 7.2705in;  cropleft "0";  croptop "1";  cropright "1";
%cropbottom "0";  filename 'Harperfig3com.eps';file-properties "XNPEU";}}}%
%BeginExpansion
\begin{figure}
[ptb]
\begin{center}
\includegraphics[
height=2.239in,
width=3.186in
]%
{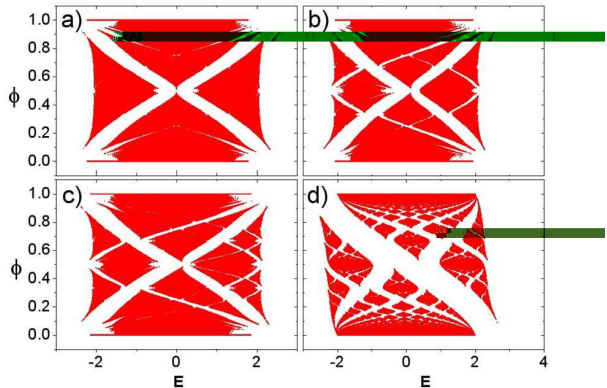}%
\caption{The energy spectrum as a function of the parameter $\phi$ for $\delta
V=1$ using a) one harmonic (corresponding to the Harper model at
$\lambda=0.2967)$, b) two harmonics, c) three harmonics, and d) full Fibonacci
potential.}%
\end{center}
\end{figure}
%EndExpansion

In Figure 3, a similar set of figures presents what happens when $\delta V=1$.
This corresponds to $\lambda=\sin(\pi\tau^{-1})\approx0.2967$ in the pure
Harper case, and the spectrum is continuous. Figure 3b) and 3c) are the cases
with two and three harmonics. Finally Fig. 3d) shows the case of the pure
Fibonacci chain, which is a beautiful fractal. Since it is known that the FC
presents a singular continuous spectrum \cite{Suto}, it is open the question
for which harmonic there is the transition from one type of spectrum to the other.

In all of the previous cases, $\phi$ was studied between $0$ and $1$ since
using the following identity,%
\[
\cos(2\pi x)=\cos(2\pi(\left\lfloor x\right\rfloor +\{x\}))=\cos(2\pi\{x\}),
\]
we have that $\cos(2\pi\phi sn+k)=\cos(2\pi\{\phi\}sn+k)$. Thus, the problem
has periodicity $1$ in $\phi$, since it depends on $\phi$ only through
$\{\phi\}$. For rational $\phi$ of the form $P/Q$, with $P$ and $Q$ integers,
this means $P<Q$.

\section{\bigskip Discussion}

In this section we will discuss the main features that arises from the
previous figures. The spectral properties of the pure Hosftadter butterfly has
been discussed by many others using diverse techniques \cite{Hofstadter}%
\cite{Aubry}\cite{Thouless}, but in order to understand the transition between
butterflies, here we will explain the main features using the structure in
reciprocal space of the potential. Let us first study the tight-binding
equation in the usual approach. We propose that the wave-function can be
written as \cite{Aubry},%
\[
\psi_{n}=e^{ikn}%
%TCIMACRO{\dsum \limits_{m=-\infty}^{\infty}}%
%BeginExpansion
{\displaystyle\sum\limits_{m=-\infty}^{\infty}}
%EndExpansion
d_{m}e^{im(2\pi\phi n+v)},
\]
where $d_{m}$ is the $m$ Fourier component of the wave function. If we
introduce this solution into Eq. (\ref{tb}), when $S$ harmonics are present in
the potential, the following set of equations are obtained,%
\begin{align}
\left(  E^{\ast}-2\delta V^{\ast}\cos(2\pi\phi m+k)\right)  d_{m}  &
=d_{m+1}+d_{m-1}+\label{reciprocal}\\
&
%TCIMACRO{\dsum \limits_{s=2}^{S}}%
%BeginExpansion
{\displaystyle\sum\limits_{s=2}^{S}}
%EndExpansion
\widetilde{V}^{\ast}(s)\left(  d_{m-s}+d_{m+s}\right)  ,
\end{align}
where the parameters are defined as $E^{\ast}=2E/(V(1)\delta V)$, $\delta
V^{\ast}=2/(V(1)\delta V),$ and$,$%
\[
\widetilde{V}^{\ast}(s)=\frac{\widetilde{V}(s)}{V(1)}=\frac{1}{s}\frac
{\sin(\pi s\phi)}{\sin(\pi\phi)},
\]
The other parameter is $k=\upsilon$ for the pure Harper equation, and
$k=v=\pi\phi$ for the general case. If only the first harmonic of the FC is
used, the equation shows that a Harper equation in the reciprocal space is
also a Harper equation with a renormalized set of parameters \cite{Thouless}.
In the case of Fibonacci, Eq. (\ref{reciprocal}) proves that the
Schr\"{o}edinger equation in reciprocal space has a different form, since each
site interacts with infinite many others. The interaction between Fourier
components decreases as $1/s$. It is a long range interaction that can be
thought as a modulating field. Observe that in the transformation between the
Hofstadter and the Fibonacci butterfly, the limitation to $S$ harmonics in the
potential, is equivalent in Fourier space to a cut-off of range $S$ of the interaction.

When $\phi$ is a rational number, say $P/Q$, the potential with any number of
harmonics has periodicity $Q$, as has been discussed in the previous section.
The corresponding wave functions are given by,%
\begin{equation}
\psi_{n}=e^{ikn}%
%TCIMACRO{\dsum \limits_{m=0}^{Q}}%
%BeginExpansion
{\displaystyle\sum\limits_{m=0}^{Q}}
%EndExpansion
d_{m}e^{im(2\pi n\frac{P}{Q}+\pi\frac{P}{Q})}, \label{waveaprox}%
\end{equation}
where the -$\pi/Q<k\leq\pi/Q$. The only difference between the Harper and the
Fibonacci case, is in the values of the $Q$ coefficients $d_{m}$. Since the
set $d_{m}$ can be obtained in the Harper case using quantum groups
\cite{Wiegmann}. It is worthwhile mentioning that previous efforts in building
a perturbation theory to tackle the proprieties of quasiperiodic Hamiltonians
have failed due to the small divisor problem, that shows up in perturbation
expansions. The present approach does not have such problem, since a
quasiperiodic solution can be built from a perturbation of a solution that is
already quasiperiodic.

Also, we can show that an effective potential can be written in the reciprocal
space for periodic approximants of the Fibonacci chain. According to Eq.
(\ref{waveaprox}), the solution must have periodicity $Q$. with $l$ an
integer. Such result can also be obtained from $V(n)$ when $\phi$ is the
rational $P/Q$. In this case, the factor $\cos\left(  2\pi s\phi n\right)  $
in Eq.(\ref{pot}) is repeated for the harmonics $s^{\prime}$ that has the form
$s+lQ$ where $l$ is a positive integer$.$ By grouping all the harmonics module
$Q$, the potential is written as,%
\begin{equation}
V(n)=\overline{V}+2\delta V%
%TCIMACRO{\dsum \limits_{s=1}^{Q}}%
%BeginExpansion
{\displaystyle\sum\limits_{s=1}^{Q}}
%EndExpansion
\widetilde{V}_{P/Q}(s)\cos\left(  2\pi s\frac{P}{Q}n\right)  ,
\end{equation}
where$\widetilde{\text{ }V}_{P/Q}(s)$ is an effective potential,%
\begin{equation}
\widetilde{V}_{P/Q}(s)=\frac{Q\sin(\pi sP/Q)}{P}\left(
%TCIMACRO{\dsum \limits_{l=0}^{\infty}}%
%BeginExpansion
{\displaystyle\sum\limits_{l=0}^{\infty}}
%EndExpansion
\frac{(-1)^{lP}}{s+lQ}\right)  . \label{Veffective}%
\end{equation}
The equation in reciprocal space is reduced as follows,
\begin{align}
\left(  E^{\prime}-2\delta V^{\prime}\cos(2\pi\phi m+k)\right)  d_{m}  &
=\label{qharmonicsrec}\\
&
%TCIMACRO{\dsum \limits_{s=1}^{Q}}%
%BeginExpansion
{\displaystyle\sum\limits_{s=1}^{Q}}
%EndExpansion
\widetilde{V}_{P/Q}(s)\left(  d_{m-s}+d_{m+s}\right)  ,
\end{align}
which shows that for a periodic approximant, the range of the interaction in
reciprocal space is $Q$.

Concerning the spectrum, the band edges are obtained from Eq.
(\ref{reciprocal}) when $k=0$ and $k=\pi/Q$. Instead of following this path,
we will look at how the structure of the potential in the\textit{ lattice
reciprocal space} determines the spectrum for $\delta V\ll1$. A much more
physical insight is obtained in this way. This approach is different from the
one realized in others works \cite{Aubry}\cite{Thouless}, since usually the
potential is projected in the base $e^{im(2\pi\phi n+v)}.$ Here we will
project into the reciprocal "vectors" of the lattice ($G$) The main idea is
that for a one dimensional crystal, it is known that each reciprocal "vector"
$G$ with component of the potential ($\widetilde{V}(G)),$opens a gap of size,%
\[
\Delta_{G}\approx2\left\Vert \widetilde{V}(G)\right\Vert ,
\]
at reciprocal vectors $q=G/2,G,3G/2,...$ It is possible to follow the opening
of the gaps by the effect of $\widetilde{V}(G).$ The reciprocal components
are,%
\[
\widetilde{V}(G)=\frac{1}{\sqrt{N}}%
%TCIMACRO{\dsum \limits_{n=0}^{N-1}}%
%BeginExpansion
{\displaystyle\sum\limits_{n=0}^{N-1}}
%EndExpansion
V(n)e^{-iGn},
\]
where $G$ can be chosen among the wave vectors $q=2\pi t/N$, with
$t=0,...,N-1$, in a lattice with $N$ sites, were periodic boundary conditions
are used. The first Brioullin zone is the interval $-\pi\leq q<\pi$, although
to simplify the algebra we take for the moment, $q$ between $0$ and $2\pi$.
Consider first the Fourier component of the Harper potential ($\widetilde
{V}_{H}(G)$) for a given parameter $\phi$,%
\begin{equation}
\widetilde{V}_{H}(G)\equiv\widetilde{V}_{H}(t)=\frac{2\lambda}{\sqrt{N}}%
%TCIMACRO{\dsum \limits_{n=0}^{N-1}}%
%BeginExpansion
{\displaystyle\sum\limits_{n=0}^{N-1}}
%EndExpansion
\left(  \frac{e^{i2\pi\phi n}+e^{-i2\pi\phi n}}{2}\right)  e^{-i2\pi tn/N}.
\label{vncomponent}%
\end{equation}
If ($\phi-t/N)$ is an integer, the problem is almost solved, because
$\widetilde{V}(t)=\lambda\delta(\phi-t/N)$. This happens whenever $\phi=P/Q$
and,%
\[
\frac{P}{Q}=\frac{m}{N},
\]
so when $N$ is chosen as a multiple of $Q,$ as for example $N=lQ$ with
$l=0,1,2,...$, then the harmonic $m=P$ is the only one that has a
contribution, and the problem is solved as a simple cosine potential.

This solution is very simple compared with other complex approaches, since
what most of the people do is fix $N$ and then a sweep of the $\phi$ is made.
In other words, the Hosftadter butterfly is built for a fixed $N$. However,
the present approach shows that if we fix a rational $\phi$ and move $N$ for
each $\phi$ until $N=lQ$, the potential is much more tractable. In the limit
of big $l$, $N\gg Q$ and there are so many vectors $k$ that the condition is
very easily achieved; in other words, there is a continuum of $q$ so
$e^{i2\pi\phi n}$ has a "possible periodicity" of the system. The Fibonacci
case is a little bit more difficult, because the potential is a sum of
cosines, but from Eq. (\ref{Veffective}) is clear that the components in
Fourier space, that we denote by $\widetilde{V}_{F}(G)$, are given by
$\widetilde{V}_{F}(G)=\widetilde{V}_{P/Q}(t)$.

The problems arise when $\phi$ is an irrational or $N$ is not a multiple $Q$,
since the previous trick is not valid. Eventually, the sum in Eq.
(\ref{vncomponent}) can be made when $\phi-t/N$ is not an integer to give,%
\[
\widetilde{V}_{H}(G)\equiv\widetilde{V}_{H}(t)=\frac{\lambda}{\sqrt{N}}\left(
\frac{1-e^{i2\pi\phi N}}{1-e^{i2\pi(\phi-t/N)}}+\frac{1-e^{-i2\pi\phi N}%
}{1-e^{-i2\pi(\phi+t/N)}}\right)  .
\]
The corresponding norm is,%
\begin{equation}
\left\Vert \widetilde{V}_{H}(t)\right\Vert ^{2}=\frac{\lambda^{2}}{N}\frac
{A}{B},\label{Vh}%
\end{equation}
where,%
\begin{align*}
A &  =(1-\cos2\pi\phi N)^{2}+\left[  \cos2\pi\phi-\cos2\pi\phi(N-1)\right]
^{2}\\
&  -2\cos(2\pi\phi t/N)\left[  1-\cos2\pi\phi/N\right]  \left[  \cos2\pi
\phi-\cos2\pi\phi(N-1)\right]  ,
\end{align*}
and,%
\[
B=\left[  1-\cos\left(  2\pi\left(  \phi-t/N\right)  \right)  \right]  \left[
1-\cos2\pi\left(  \phi+t/N\right)  \right]  .
\]
Notice how $\widetilde{V}_{H}(t)\rightarrow\lambda\delta(\phi-t/N)$ as
$\phi\rightarrow t/N$. The gaps widths depend upon these components, and the
maximum of $\widetilde{V}_{H}\left(  G\right)  $ occurs when $G\approx\pm\phi
$. Now let us propose the solution,
\[
\psi_{n}=%
%TCIMACRO{\dsum \limits_{q=-\infty}^{\infty}}%
%BeginExpansion
{\displaystyle\sum\limits_{q=-\infty}^{\infty}}
%EndExpansion
c_{q}e^{iqn},
\]
for Eq. (\ref{tb}). The resulting equation in reciprocal space is,%
\[
\left(  E-2\cos q\right)  c_{q}=%
%TCIMACRO{\dsum \limits_{G}}%
%BeginExpansion
{\displaystyle\sum\limits_{G}}
%EndExpansion
\widetilde{V}_{H}(G)c_{q-G}%
\]
Since the main contribution to $\widetilde{V}_{H}(G)$ comes from $G\approx
\pm\phi$, a mixing of wave vectors $G\approx\phi$ and $G\approx-\phi$ occurs
at $q=\pm\phi/2$. Using perturbation theory, this means that for $\lambda\ll
1$, the main gaps are open around$,$%
\[
E\approx\pm\lambda\cos(2\pi(\phi/2))=\pm\lambda\cos(\pi\phi).
\]
Figure 4 compares this prediction with the Hofstadter butterfly, showing an
excellent agreement. Other band gaps are obtained at $q=\pm r\phi/2$ with
$r\in%
%TCIMACRO{\U{2115} }%
%BeginExpansion
\mathbb{N}
%EndExpansion
$. The general $p$th-order perturbation term is of the form,%
\[
V_{q}=V(q_{0}-q_{1})V(q_{1}-q_{2})...V(q_{N-1}-q_{N})
\]
where $q_{t}$ labels the vector $2\pi t/N$. For small $\lambda,$ gaps will be
open at,
\[
E\approx\lambda^{r}\cos(\pi\phi r),\qquad
\]
These cosine branches are also plotted in Figure 4, compared with the
Hofstadter butterfly, showing that the basic structure of the spectrum is
determined by these branches. When $\lambda$ is near one, around each gap
there are many wave vectors that mix together, so the present approximation
breaks out.%
%TCIMACRO{\FRAME{ftbpFU}{2.9032in}{2.0384in}{0pt}{\Qcb{General structure of
%\ all the obtained spectra. The Hofstadter butterfly is compared with the
%cosine branches given by $E=\lambda\cos(\phi m)$ for $m=1,2$ and $3$ for
%$\lambda=0.2967$.}}{}{harperfig4com.eps}%
%width 2.9032in;  height 2.0384in;  depth 0pt;  original-width 8.9015in;
%original-height 6.2405in;  cropleft "0";  croptop "1";  cropright "1";
%cropbottom "0";  filename 'Harperfig4com.eps';file-properties "XNPEU";}}}%
%BeginExpansion
\begin{figure}
[ptb]
\begin{center}
\includegraphics[
height=2.0384in,
width=2.9032in
]%
{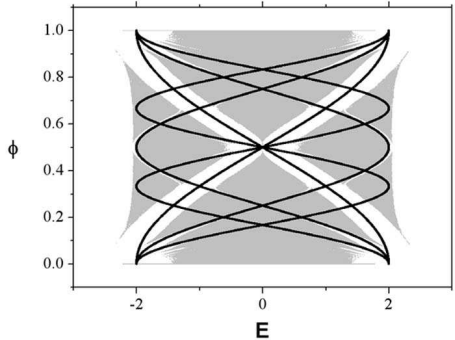}%
\caption{General structure of \ all the obtained spectra. The Hofstadter
butterfly is compared with the cosine branches given by $E=\lambda\cos(\phi
m)$ for $m=1,2$ and $3$ for $\lambda=0.2967$.}%
\end{center}
\end{figure}
%EndExpansion

A detailed observation of the Fibonacci spectrum, shows that the main effect
upon the spectrum is an asymmetry of the cosine branches, and the different
gap sizes around these branches, due to the mixing of different modulating
harmonics. This is very clear in Figures 3 b) and 3 c). The same analysis
performed to the Harper equation is valid for these cases, specially for low
$S$. . The equivalent rough approximation for the FC is,%
\[
E\approx\pm\left(  \delta V%
%TCIMACRO{\dsum \limits_{s=1}^{\infty}}%
%BeginExpansion
{\displaystyle\sum\limits_{s=1}^{\infty}}
%EndExpansion
\frac{\sin(\pi s\phi)}{\pi s}\right)  \cos(\pi\phi),
\]

What is behind the process of gap opening, is a self-similar folding of the
bands in the reciprocal space, as revealed by following the sequence of
approximants for $\phi$ in all the obtained spectra. For $\phi=0$, the
spectrum is continuous and goes from -$2\delta V$ to $2\delta V.$ The next
most simple spectrum corresponds to $\phi=1/2$, which gives two bands. If the
lattice has period $2$, then the first Brillouin zone of the zero approximant
($\phi=0),$ is folded around half the original Brioullin zone limit,
$k=\pi/2.$ As usual, an energy gap of size $\Delta_{\mathbf{G}}\approx
2\left\vert \widetilde{V}(\mathbf{G})\right\vert $ is open in the zone
boundary due to a mixing of waves with reciprocal vectors $\mathbf{G}/2$ and
-$\mathbf{G}/2$. In this case, $\mathbf{G=}\pi/2$ and $\widetilde
{V}(\mathbf{G})=2\lambda\mathbf{.}$ For the Harper equation we get,
\[
\Delta_{\mathbf{G}}=4\lambda,
\]
In the Harper equation, there is symmetry around $E=0$, thus the bands limits
are $E=\pm2\lambda$. For the pure FC first approximant ($\phi=1/2$),
$\widetilde{V}(\mathbf{G})$ produces a different gap, and the central gap
limits are,%

\[
\Delta_{\mathbf{G}}=\pm\delta V.
\]
This gap is clearly in the horizontal lines at $\phi=1/2$ in Fig. 3 d). The
process of folding in reciprocal space can be repeated in a similar way for
other rationals like $\phi=2/3$ and $\phi=1/3.$In this case, the periodicity
is $3$, and the folding around the first Brioullin zone limit occurs at
$k=\pi/3.$Three bands are produced in this case, and the gaps are centered at
$\pm\lambda\cos(\pi2/3)$ and $\pm\lambda\cos(\pi/3)$. The process is repeated
for other approximants..

\section{Localization properties using correlators}

As an example of the utility of having an expansion of the Fibonacci
potential, we show how to obtain certain localization properties from the
potential correlators. This can serve to understand how the addition of
harmonics leads to different localization properties. For $\lambda=1$ the
eigenstates of the Harper equation are critical, while they are localized for
$\lambda>1$ and $\lambda<1$. This comes from Eq.(\ref{reciprocal}). Therein,
if the square of the components is finite,%
\[%
%TCIMACRO{\dsum \limits_{m=-\infty}^{\infty}}%
%BeginExpansion
{\displaystyle\sum\limits_{m=-\infty}^{\infty}}
%EndExpansion
\left\vert d_{m}\right\vert ^{2}<\infty
\]
and the corresponding wave-function is non-localized. In the pure Harper
equation, when $\lambda\rightarrow\infty$, the solutions in real space are
localized wave-functions, since it corresponds to the dual of a solution
$\lambda\rightarrow0$ in reciprocal space, which are known to be extended.

For the Fibonacci chain, it is known that for all values of $\lambda$ the
eigenstates are critical \cite{NaumisJPCM}. An interesting question is how
many harmonics are needed to produce this transition. Although this requires a
complete study, let us show how this problem can be tackled. To measure
localization in one dimension, the Lyapunov exponents are used ($L^{-1}$).
They are the inverse of the localization length ($L$) for an exponential
localized state. It has been shown that for one dimensional Hamiltonians, the
localization properties depend upon the pair correlation of the potential
\cite{Izrailev} when $V(n)\ll1$. Notice that this is the most interesting
limit for quasiperiodic potentials, since for $V(n)\gg1$ there are already
many methods to treat the spectrum and the eigenfunctions \cite{Niu}. For a
one dimensional Hamiltonian \cite{Izrailev},%
\[
L^{-1}=\frac{\epsilon_{0}\varphi(\mu)}{8\sin^{2}(\mu)};\qquad\varphi(\mu)=1+2%
%TCIMACRO{\dsum \limits_{k=1}^{\infty}}%
%BeginExpansion
{\displaystyle\sum\limits_{k=1}^{\infty}}
%EndExpansion
\xi(k)\cos(2\mu k)
\]

Here, the function $\varphi(\mu)$ is given by the Fourier series with the
coefficients $\xi(k)$ which is the correlator of the site potential $V(n)$.
The correlator $\xi(k)$ is defined as,%
\[
\left\langle V(n)V(n+k)\right\rangle =\epsilon_{0}^{2}\xi(k),\qquad
\left\langle V(n)V(n)\right\rangle =\epsilon_{0}^{2}%
\]

For the Harper equation, $\xi(k)=\cos(2\pi\phi k),$ which gives $\varphi
(\mu)=0$ and thus all states are non exponentially localized. In the case of a
FC, by using Eq. (\ref{pot}), the correlator can be written as,%
\[
\xi(k)=4\delta V%
%TCIMACRO{\dsum \limits_{s=1}^{S}}%
%BeginExpansion
{\displaystyle\sum\limits_{s=1}^{S}}
%EndExpansion
\frac{\sin^{2}(\pi s\phi)}{(s\phi)^{2}}\cos(2\pi s\phi k)
\]
which also gives $\varphi(\mu)=0$ for all $S$. Thus, we have proved that all
states are non exponentially localized for any number of harmonics. The states
are extended or critical. An interesting question that remains to be answered,
is how the transition from extended to critical states is achieved as the
number of harmonics is increased from Harper to Fibonacci. To solve this
question, an expression for the scaling exponents in terms of the correlators
is needed \cite{Naumisloc}.

\section{Conclusions}

In the present article, we have shown that the Fibonacci potential can be
written as a sum of Harper potentials. As a consequence, one can follow the
evolution of the spectral types as a function of the number of harmonics. In
particular, a butterfly similar to the Hofstadter case is found for the
Fibonacci. The corresponding spectrum is a fractal object. The Fourier
components of the potential provides a simple explanation for the main
features of the spectra. We also show how the development of the potential
leads to interesting questions, as for example, at which harmonic the spectral
type is changed. This is equivalent to ask for which harmonic the
eigenfunctions of the Fibonacci case become critical. The present approach
also leads to the possibility of building analytical solutions for a FC. We
hope that other researchers will answer some of these interesting questions in
the nearby future.

We would like to thank DGAPA-UNAM projects IN-117806 and IN-111906-3 for
financial help.


\begin{thebibliography}{99}                                                                                               %


\bibitem {Blech}D. Shechtman, I. Blech, D. Gratias, J.W. Cahn, Phys. Rev.
Lett. 53, 1951 (1984).

\bibitem {Bambi}P. Tong, B. Li, B. Hu, Phys. Rev. Lett. 88, 046804 (2002) .

\bibitem {Hofstadter}D.R. Hofstadter, Phys. Rev. B 14, 2239 (1976) .

\bibitem {Aubry}S. Aubry, G. Andre, in: Colloquium on Group Theoretical
Methods In Physics, Preprint, (1979). Reprinted in: The physics of
quasicrystals, ed. by P. Steinhardt and Stellan Ostlund, World Scientific, Singapore,1987.

\bibitem {Ketmerick}R. Ketzmerick, K. Kruse, S. Kraut, T. Geisel, Phys. Rev.
Lett. 79, 1959 (1997).

\bibitem {Machida}K. Machida, M. Fujita, Phys. Rev. B 34, 7367 (1986) .

\bibitem {Geisel}T. Geisel, R. Ketzmerick, G. Petschel, Phys. Rev. Lett. 66,
1651 (1991).

\bibitem {Wiegmann}P.B. Wiegmann, A.V. Zabrodin Phys. Rev. Lett. 72,
1890--1893 (1994).

\bibitem {Koshino}M. Koshino, H. \ Aoki, T. Osada, K. Kuroki, S. Kagoshima,
Phys. Rev. B65, 045310 (2002).

\bibitem {Czajka}K. Czajka, A. Gorczyca, M.M. Maska, M. Mierzejewski, Phys.
Rev. B74, 125116 (2006).

\bibitem {Kohmoto}Ch. Tang, M. Kohmoto, Phys. Rev. B 34, 2041 (1986).

\bibitem {Thiel}P. Thiel, and J.M. Dubois, Nature 406, 571 (2000).

\bibitem {Vedemenko}E.Y. Vedmendenko, U. Grimm, R. Wiesendanger, Phys. Rev.
Lett. 93, 076407 (2004) .

\bibitem {Moras}P. Moras, et. al., Phys. Rev. Lett. 96, 156401 (2006) .

\bibitem {Maciarep}E. Macia, Rep. Prog. Phys. 69, 397 (2006).

\bibitem {Kittel}Ch. Kittel, Introduction to Solid-State Physics, 7th. ed.,
Wiley, New York, 1996.

\bibitem {Zallen}Zallen, The physics of amorphous solids, J. Wiley \& Sons,
New York, 1983.

\bibitem {Suto}A. S\"{u}t\"{o}, in Beyond Quasicrystals, ed. by F.Axel and D.
Gratias, Les Editions de Physique, France, 1994.

\bibitem {NaumisPenrose}G.G. Naumis, R.A. Barrio, Ch. Wang, Phys. Rev. B 50,
9834 (1994).

\bibitem {NaumisJPC}G.G. Naumis, J. Phys.: Condens. Matter 11, 7143 (1999).

\bibitem {Dominguez}E. Maci\'{a}, F. Dom\'{\i}nguez-Adame, Phys. Rev. Lett.
79, 5301 (1991).

\bibitem {Fujiwara}T. Fujiwara, M. Kohmoto, T. Tokihiro, Phys. Rev. B 40, 7413 (1989).

\bibitem {Macia}E. Macia, Phys. Rev. B 60, 10 032 (1999).

\bibitem {Naumisphason}G.G. Naumis, Ch. Wang, M.F. Thorpe, R.A. Barrio, Phys.
Rev. B 59, 14 302 (1999).

\bibitem {Yeong}Gi-Yeong Oh., Phys. Rev. B60, 1939 (1999).

\bibitem {NaumisJPCM}G.G. Naumis, J. of Phys.: Cond. Matter. 15, 5969 (2003).

\bibitem {Thouless}D.J. Thouless, Phys. Rev. B 28, 4272 (1983).

\bibitem {Izrailev}F.M. Izrailev, A.A. Krokhin, Phys. Rev. Lett. 82, 4062(1999).

\bibitem {Niu}Q. Niu, T. Odagaki, J.L Birman, Phys. Rev. B 42, 10329 (1990).

\bibitem {Naumisloc}G.G. Naumis, to be published.
\end{thebibliography}
\end{document}